\documentclass[12pt]{iopart}

\usepackage{times}
\usepackage{graphicx}
\usepackage{natbib}

\newcommand{\msigmahae}{$M_{\bullet}$--$\sigma_*$\ }
\newcommand{\lahae}{\mathrel{\hbox{\rlap{\hbox{\lower4pt\hbox{$\sim$}}}\hbox{$<$}}}}
\newcommand{\gahae}{\mathrel{\hbox{\rlap{\hbox{\lower4pt\hbox{$\sim$}}}\hbox{$>$}}}}
\newcommand{\dhae}{{\rm d}}
\newcommand{\kmshae}{\mbox{km\,s$^{-1}$}}
\newcommand{\yrhae}{\mbox{yr}}
\newcommand{\Msolhae}{\mbox{M$_\odot$}}

\begin{document}

\title[]{Hierarchical build-up of galactic bulges and the merging 
rate of supermassive binary black holes}

\author{Martin G. Haehnelt\dag\   
\footnote[3]{haehnelt@ast.cam.ac.uk}
}

\address{\dag\ Institute of Astronomy, Madingley Road, CB3 0HA,
  Cambridge, UK}

\begin{abstract}
The hierarchical build-up of galactic bulges should lead to the
build-up of present-day supermassive black holes by a mixture of 
gas accretion and merging of supermassive black holes. The tight  
relation between black hole mass and stellar velocity 
dispersion is thereby a strong argument that the supermassive 
black holes in merging galactic bulges do indeed merge. Otherwise 
the ejection of supermassive black holes by gravitational slingshot 
would lead to excessive scatter in this relation. At high redshift
the coalescence of massive black hole binaries is likely to be driven 
by the accretion of gas in the major mergers signposted by optically 
bright QSO activity. If massive black holes only form 
efficiently by direct collapse of gas in deep galactic potential 
wells with $v_{\rm c}  \gahae 100 \kmshae$ as postulated  in the model of 
Kauffmann \& Haehnelt (2000) LISA expects  to see event rates from 
the merging of massive binary black holes of about 0.1-1 ${\rm}
\yrhae^{-1}$  spread over the redshift range 
$0\le z\le 5$. If, however, the hierarchical build-up of supermassive 
black holes extends to pre-galactic structures with significantly 
shallower potential wells  event rates may be as high   as 10-100
$\yrhae^{-1}$ and will be dominated by events from redshift $z \gahae 5$.

\end{abstract}



\section{Hierarchical build-up of supermassive black holes and
galactic bulges} 

According to the standard paradigm of structure formation in the 
Universe, galaxies merge frequently as their dark matter halos 
assemble. This process has been modeled extensively using 
Monte-Carlo realizations which include simple prescriptions  
to describe gas cooling, star formation, supernova feedback and 
merging rates of galaxies (see e.g. Kauffmann et al. 1999 
for a recent account).  Kauffmann \& Haehnelt (KH2000) introduced a 
``unified'' model for the evolution of galaxies and quasars in a 
cold dark matter (CDM) dominated Universe. In the model of KH2000, 
spheroids form when two galaxies of comparable mass merge.  
The resulting gas accretion onto the merging black holes leads to QSO 
activity which lasts  for a few times $10^{7}$ yr. The model of KH2000 
is able to reproduce  a wide variety of galaxy and QSO properties like the 
galaxy/QSO luminosity function and its evolution with redshift, host 
galaxy luminosities, the $M_{\bullet}$--$\sigma_*$\ relation (
Ferrarese\& Merritt 2000; Gebhardt et al. 2000; Haehnelt \& 
Kauffmann 2000),  and the clustering properties of galaxies and 
QSOs (Kauffmann \& Haehnelt 2002).  The  wealth of observational 
constraints gives some confidence that the model gives 
a  realistic account of  the expected merging rate of galactic bulges 
and the gas  accretion history of the central supermassive black holes 
in typical  nearby bright galaxies despite the simplicity with which 
relevant physical and dynamical processes are 
modeled.

\begin{figure}
\centerline{\resizebox{\columnwidth}{!}
{\includegraphics[angle=0,scale=0.9,clip]{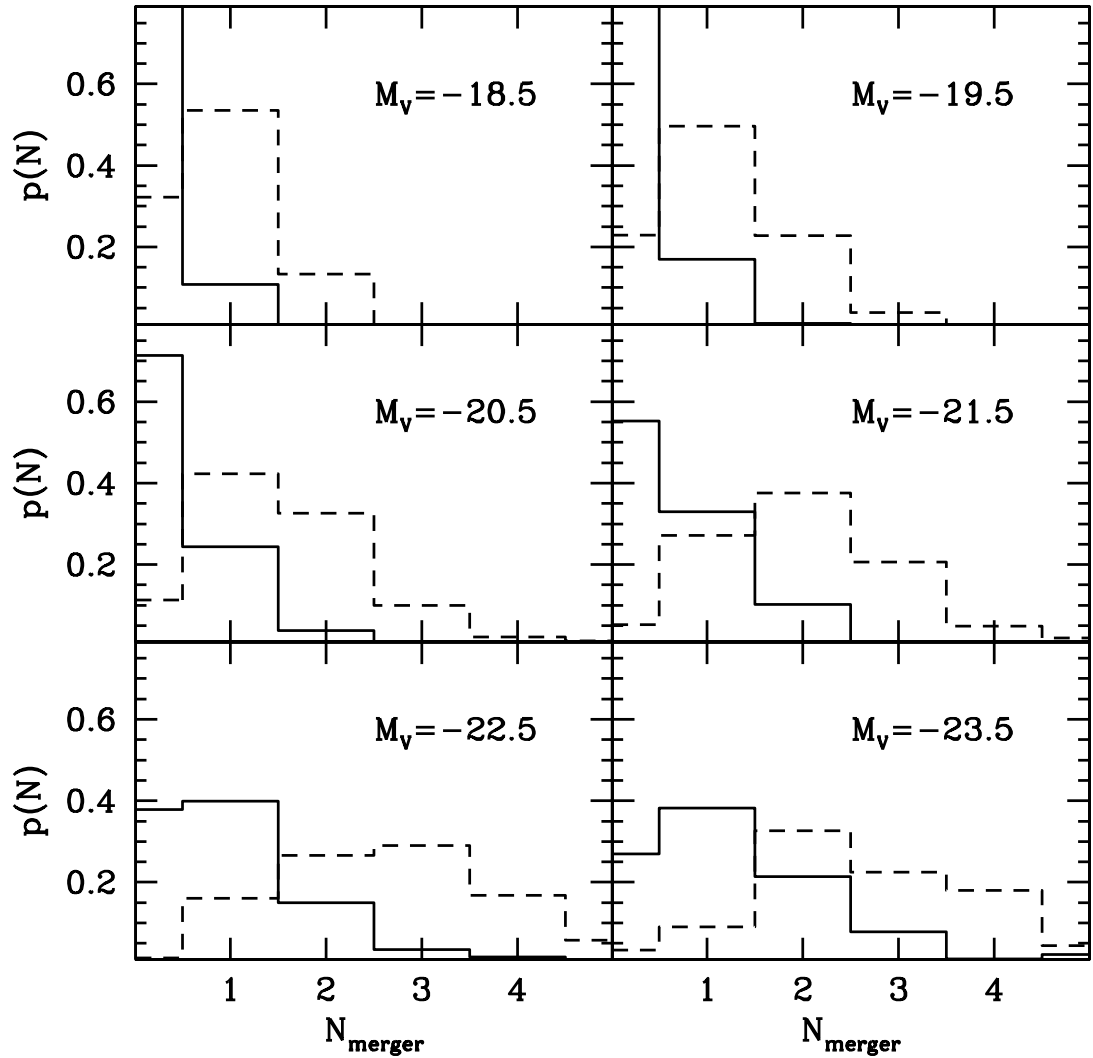}}}
\caption[]{The probability distribution 
of the number number of mergers expected to lead to the formation 
of massive binary black holes with mass ratios 
$>0.3$ in galactic bulges with different  V-band luminosities. 
The dashed curves show the total number, while the solid curves 
show the number after the last major  gas accretion event (Haehnelt\&
Kauffmann 2002).}
\label{fig:fig4}
\end{figure}

\section{Supermassive binaries in present-day galactic bulges} 

The frequent merging of galaxies which each contain one or more black 
holes will lead to the occurrence of multiple black holes. 
When two galaxies merge, the smaller galaxy will sink to the centre of the 
merger remnant  because of dynamical friction. The outer regions of 
the infalling galaxy will be gradually tidally stripped in the process. 
If two galaxies with roughly equal mass merge, 
a binary black hole will form within a few dynamical times 
(e.g. Milosavljevi\'c \& Merritt 2001). The subsequent evolution of the 
supermassive binary once the binary is hard ({\it i.e.} its orbital 
velocity is comparable to the velocity dispersion of the stars)
has been first discussed by Begelman, Blandford \& Rees (1980). 
The binary is expected to harden either by gravitational 
sling-shot ejection of stars ({\it e.g.}  Quinlan 1996) 
or by the accretion of gas onto the binary system 
(Armitage \& Natarajan 2002). The timescale for the ejection 
of stars to cause the binary to harden may  exceed the 
Hubble time in bright galaxies (Yu 2002). This raises the possibility   
of triple interactions of supermassive black holes which may lead 
to the ejection of black holes  from the centre by gravitational 
slingshot (Saslaw, Valtonen \& Aarseth 1974).
The frequent merging 
predicted by hierarchical galaxy formation models is then 
inconsistent with the observed tight $M_{\bullet}$--$\sigma_*$\ 
relation unless the gas accretion during phases  of QSO activity
merges black holes efficiently. 
\begin{figure}
\centerline{\resizebox{\columnwidth}{!}{\includegraphics[angle=0,scale=0.9,clip]{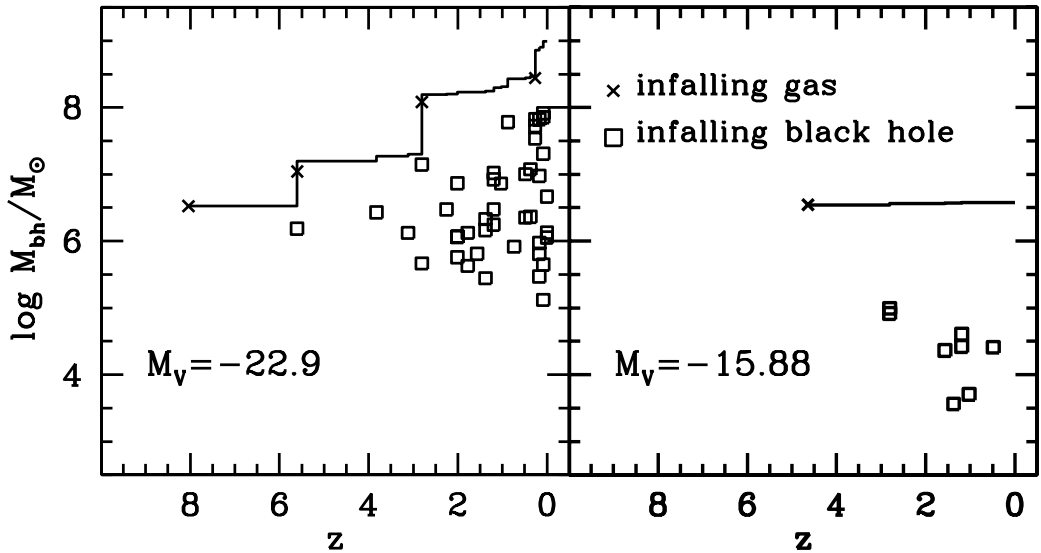}}}
\caption[]{Typical accretion/merging history of the central black hole 
in a bright (left) and faint (right) galactic bulge in the model of 
Kauffmann \& Haehnelt (2000).  
Crosses denote gas infall during major galaxy mergers   
and squares denote the infall of black holes.}
\label{fig:fig2}
\end{figure}
The dashed lines in Fig. 1 shows the predicted number 
of major mergers ($m_{\rm sec}/m_{\rm prim}>0.3$). 
Typical bright elliptical galaxies are expected to undergo 
several major mergers. If the merging timescale 
of supermassive binary black holes is longer than                
the Hubble time,  a binary should be ejected in 
a large fraction of bright elliptical galaxies.
This would appear to conflict with the fact that black holes are 
observed in all nearby bright elliptical galaxies  and with the  
tightness of the observed \msigmahae relation (Ferrarese\& Merritt 2000,
Gebhardt et al. 2000). 
The solid curve shows instead the predicted number of major mergers
after the last accretion event in which the mass     
of accreted gas exceeded the sum of the masses of the two black holes 
in the binary system. The median number of mergers since this 
event is significantly smaller and ranges from zero in faint galaxies 
to one in bright galaxies. 
If the supermassive  black holes do indeed merge during gas-rich
accretion events, the  fraction of elliptical galaxies containing 
large mass ratio binary black holes 
will not be larger than 10\%in faint ellipticals and 40\% in brighter objects.
The fraction of galaxies with  a third massive ``intruder''   
ranges  from 0 to 20 percent. Binary black hole ejection will 
then only occur in a small fraction  of only  the  brightest  galaxies 
(Haehnelt \& Kauffmann 2002).  
\begin{figure}
\centerline{\resizebox{\columnwidth}{!}{\includegraphics[angle=0,scale=0.9,clip]{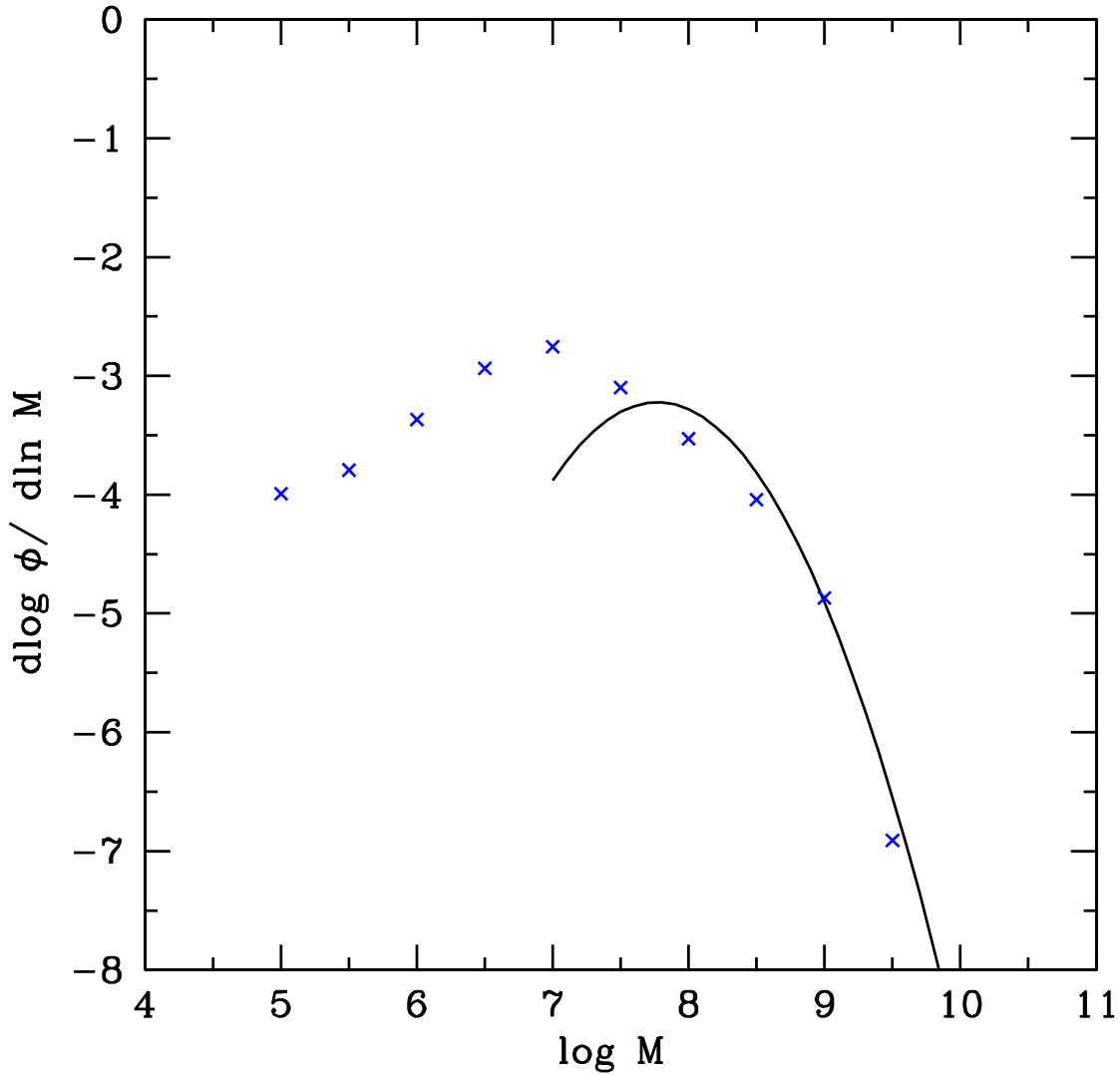}}}
\caption[]{Crosses show the present-day black hole mass function 
in the model of Haehnelt\& Kauffmann (2000). The solid curve shows the 
black hole function in early-type galaxies inferred by Yu \& Tremaine 
(2002) from the galaxy sample of Bernardi et al. (2002). Note that 
the mass function of Yu \& Tremaine does not account for black holes 
in the bulges of spiral galaxies.} 
\label{fig:fig1}
\end{figure}
Figure 2 shows  typical merging/ accretion histories of the central
black hole in a bright and faint galactic bulge. Note that the 
good agreement  between the present-day black hole mass density and
that inferred to be accreted using Soltan's argument (Soltan 1982)
does not leave much room for growth of black holes other than by 
merging and infall 
of gas during phases of optically bright QSO activity 
(Yu \& Tremaine 2002).   
The crosses in Fig. 3 show the  black hole mass function of the model of 
HK2000 while the solid curve shows the black hole mass function in early-type 
galaxies as determined by Yu \&Tremaine (2002) using the galaxy sample
of Bernardi et al. 2002. Note that the result by Yu \&Tremaine (2002)
does not take into account black holes in the bulges of spiral 
galaxies. The discrepancy with the model at the low mass end is thus
expected.  In  the model of KH2000  the efficiency of 
funneling cold gas to the centre of the galaxy is a strong function 
of the depth of the potential well and black holes initially form 
from collapse of the cold gas in potential wells with circular
velocity $v_{\rm c} \gahae 100 \kmshae$. This leads to rather massive 
effective ``seed'' black holes and the drop of the black hole mass 
function at masses $\lahae 10^{7} \Msolhae$ in Fig.3.  

\begin{figure}
\centerline{\resizebox{\columnwidth}{!}{\includegraphics[angle=0,scale=0.8,clip]{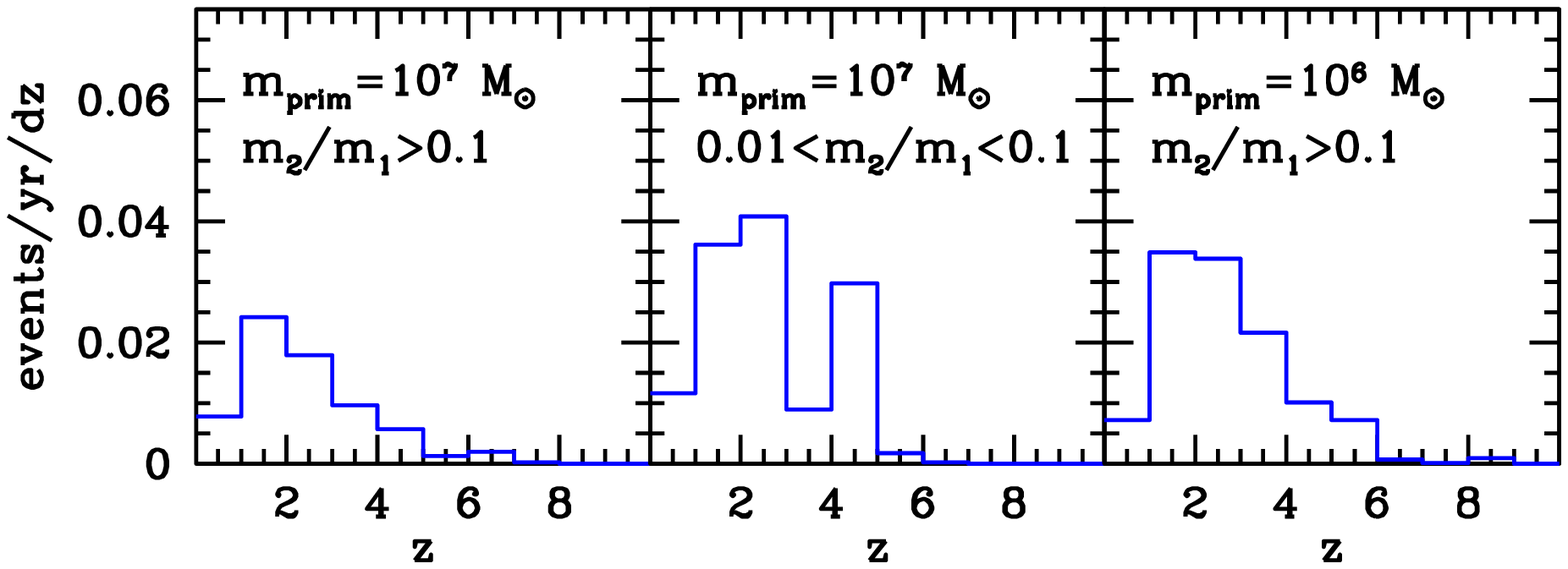}}}
\caption[]{Merger rate of galactic bulges expected to 
form  massive binary black holes for a range of primary black hole masses 
and mass ratios relevant to LISA in the model of KH2000. }      
\label{fig:fig3}
\end{figure}

\section{Merging rates of galactic bulges in the KH2000 model}

For a population of merging sources with comoving density  
of mergers per unit redshift $\dhae n_{\rm merge}/\dhae z$ 
the all-sky event rate per unit redshift can be written as 
\begin{eqnarray*}
\dhae N_{\rm merge}/\dhae z/\dhae t &=& 4\pi R(z)^2\; c\; \dhae n_{\rm merge}/ 
\dhae z\cr
&\null &\cr
&\approx& 0.08\,  (n_{\rm merge}/10^{-3} {\rm Mpc} ^{-3}) \;   r(z)^2 \;
{\rm yr}^{-1} \cr
\end{eqnarray*} 
where $R(z)=  r(z) c/H_{0}$ and $r(z)  =
 1/ \int{(\Omega_{\rm m} (1+z)^3  + (1-\Omega_{\rm m} -\lambda) (1+z)
 ^2 +  \lambda \dhae z}) $ and h=0.65 was assumed. 
The planned gravitational wave interferometer 
LISA is expected to detect black hole coalescences 
with primary black hole masses of $10^{6} - 10^{7} \Msolhae$
and mass ratios as small as 0.01 out to very large redshifts 
(e.g. Bender these proceedings).  Figure 4 shows the  merging 
rates  of galactic bulges expected to form massive 
binary black holes with the mass of the primary in this range 
for different mass ratios. Note, however, that in order to interpret 
these as potential LISA event rates one has to assume that these binaries
do neither get ``hung-up'' (Begelman, Blandford \& Rees 1980; 
Milosavljevi\'c these proceedings) nor  are ejected by gravitational 
sling-shot (Saslaw, Valtonen \& Aarseth 1974). The total predicted rate
for the mass range shown in Figure 4 is 0.3 $\yrhae^{-1}$
spread  over the redshift range $0\le z\le 5$. 
Including events with larger primary mass and smaller mass ratios 
will increase
this rate to about 1 $\yrhae^{-1}$.  Note, however, again that the  
specific assumptions made in KH2000 on how ``seed'' black holes form 
lead to a rapidly declining mass function below $10^{7} \Msolhae$.  
These assumptions were motivated by the rather large overall
efficiency with which black holes in galactic bulges have formed.  
The black hole mass is about 0.1 percent of the stellar mass. 
Currently little is known observationally about the mass function below
$10^{6} \Msolhae$. Recent claims of the detection of intermediate mass
black holes in two globular clusters are controversial
(Gerssen  et al. 2002, Gebhardt et al. 2002). 
The same holds for the interpretation of the ultraluminous 
compact X-ray sources as intermediate mass black holes
(e.g. King et al. 2001). Nevertheless the evolution 
of a dense stellar cluster in the collisional regime and Eddington 
limited accretion onto stellar mass black holes  are plausible
mechanism for the formation of a population of intermediate mass
black holes albeit almost certainly with a smaller 
efficiency (Begelman \& Rees 1978; Rees 1984; Madau \& Rees 2000, 
Portegies Zwart \& McMillan 2002). In the next section we will discuss 
simple estimates for the expected event rates of the hierarchical
build-up of supermassive black holes from stellar or intermediate 
mass seed black holes in pre-galactic structures.

\section{Build-up of black holes in hierarchical merging 
pre-galactic structures}

The typical mass of the DM halo hosting a bulge with a $10^{7} \Msolhae$
black hole should be about  $10^{12} \Msolhae$. The hierarchical 
build-up of galaxies is expected, however,  to start at much smaller masses. 
The total merging rate of these small
DM haloes is about $10\; (M_{\rm halo}/10^{11}\Msolhae))^{-1}\yrhae^{-1}$. 
If mergers of small haloes at high redshift lead to detectable black 
hole coalescences the corresponding  event rates will  be much larger 
than those discussed in the last section (Haehnelt 1994, Haehnelt 1998, Menou 
et al. 2001). 
Unfortunately we do no not have much of a handle on the mass 
function and assembly history of black holes in these haloes 
(see Volonteri et al. 2002 for a first attempt to
model these in detail). We can nevertheless try to get at least an
upper limit on  the expected event rates which is consistent
with the observed present-day black hole mass density. Yu and 
Tremaine (2002) obtain  $\rho_{\rm bh} = 2.5 \times 10^{5} 
\Msolhae {\rm Mpc}^{-3}$  with a mean black hole mass of $\sim10^{8} \Msolhae$
(see also Aller \& Richstone 2002). 
If we assume that these build up solely by equal mass mergers of black 
holes with a minimum mass $M_{\rm min}$ this would correspond to an
event rate of $200\, (M_{\rm min}/10^{5}\Msolhae)^{-1}\, \yrhae^{-1}$. 
Obviously this
is  overly optimistic as we know that black holes must have accreted 
a  considerable fraction of their mass  when their masses were 
already larger than $10^{7} \Msolhae$ in order to  produce the light emitted
by optically bright QSOs.  The usual caveat that massive black hole binaries
may not actually merge is obviously also still valid.  
A merger rate of 10\; $\yrhae^{-1}$ may  be a more realistic estimate 
of the coalescence rate in case intermediate black holes 
were indeed forming  in dark matter haloes with $v_c < 100 \kmshae$.  
Note also, that such frequent merging of black holes in low mass DM haloes
would require   a steep black hole mass function. Most of these 
events would occur at $z\gahae 5$.

\section{Conclusion}
The frequent merging of galactic bulges expected in hierarchical
models of structure formation  together
with the fact that all galactic bulges 
appear to contain supermassive black holes 
makes the formation of supermassive binary black holes
inevitable. The tightness of the observed relation between black hole
mass and stellar velocity dispersion leaves thereby little room for hung-up 
and subsequently ejected binaries giving reason for some optimism 
that these binaries do generally coalesce within a Hubble time. 
The merging  rate of galactic bulges expected to 
form  massive binary black holes in the  
range of primary black hole masses  and mass ratios 
to which LISA will be sensitive is 0.1-1 ${\rm}\yrhae^{-1}$ 
if massive black holes only form efficiently by direct collapse 
of gas in deep galactic potential
wells with $v_{\rm c}  \gahae 100 \kmshae$, but may be as large  
as  10-100 $\yrhae^{-1}$ if the hierarchical build-up of supermassive 
black holes extends to pre-galactic structures with significantly 
shallower potential wells.

\ack
I am  grateful to my collaborator Guinevere Kauffmann for permission
to present results of our joint research. 
                                                                      
\section*{References}
\begin{harvard}

\item[] Aller M.C., Richstone D.O., 2002, AJ, in press, 
astro-ph/0210573
\item[] Armitage P., Natarajan P., 2002, ApJ, 567, L9 
\item[] Begelman M., Blandford R., Rees M.J., 1980, Nature, 287, 
307 
\item[] Begelman M., Rees M.J., 1978, MNRAS, 175, 847 
307 
\item[] Bernardi M. et al, 2002, AJ in press, astro-ph/0110344 
307 
\item[] Ferrarese L., Merritt D.,  2000, ApJ, 539, L9
\item[] Gebhardt K.,  Bender R. Bower G., 
Dressler A., Faber S.M., Filipenko A.V., Green R., Grillmair C.,
Ho L.C., Kormendy J., Lauer T.R., Magorrian J., Pinkney J., 
Richstone D., \& Tremaine S.,  2000, ApJ, 539, L13  
\item[] Gebhardt K., Rich R.M., Ho L.C., 2002, ApJ, 578, L41
\item[] Gerssen J.,  van der Marel R.P.,  Gebhardt K., 
Guhathakurta P.,  Peterson R., Pryor C., 2002, astroph/0209315 
and  astroph/0210158
\item[] Haehnelt M.G., 1994, MNRAS, 269, 199  
\item[] Haehnelt M.G., 1998, in AIP conference Proceedings 456:
Second International LISA Symposium, ed. W.M. Folkner, p. 45 
\item[] Haehnelt M.G., Kauffmann G.,  2000, MNRAS, 318, L35 (HK2000) 
\item[] Haehnelt M.G., Kauffmann G.,  2002, MNRAS, 336, L61 
\item[] Kauffmann G.,  Colberg J. M., Diaferio A.,  White S.D.M.,
1999, MNRAS, 303, 188 
\item[] Kauffmann G.,  Haehnelt M.G., 2000, MNRAS, 311, 576 (KH2000) 
\item[] Kauffmann G.,  Haehnelt M.G., 2002, MNRAS, 332, 529  
\item[] King A. R., Davies M. B., Ward M. J., Fabbiano G., Elvis M., 
2001, ApJ, 552, L109
\item[] Madau P., Rees M.J., 2001, 511, L27
\item[] Menou K., Haiman Z., Narayanan V.K.. 2001, ApJ, 558,a 535
\item[] Milosavljevi\'c,  Merritt D., 2001, ApJ, 563, 34
\item[] Portegies Zwart S., McMillan S.L. W,  2002, ApJ, 576, 899
\item[] Rees M.J., 1984, ARAA, 22, 471
\item[] Saslaw W.C., Valtonen M.J., Aarseth S.J., 1974, ApJ, 190, 253 
\item[] Soltan A., 1982, MNRAS, 200, 115
\item[] Volonteri M., Haardt F., Madau, P., 2002, ApJ, in press
 astro-ph/0207276
\item[] Yu Q., Tremaine S., 2002, MNRAS, 335, 695
\item[] Yu Q., 2002, MNRAS, 331, 935

\end{harvard}

\end{document}